\documentclass[12pt]{article}
\usepackage{a4}
\usepackage{cite}
\usepackage{graphicx}
\usepackage{amssymb}
\usepackage{amsmath}
\usepackage{epsfig}
\usepackage[tmargin=1cm,bmargin=1cm,lmargin=2cm,rmargin=2cm]{geometry}

\newcommand{\qb}{\ensuremath{\overline q}}

\newcommand{\be}{\begin{equation}}
\newcommand{\bdm}{\begin{displaymath}}
\newcommand{\bea}{\begin{eqnarray}}
\newcommand{\beastar}{\begin{eqnarray*}}

\newcommand{\ee}{\end{equation}}

\newcommand{\eps}{\ensuremath{\epsilon}}

\newcommand{\edm}{\end{displaymath}}
\newcommand{\eea}{\end{eqnarray}}
\newcommand{\eeastar}{\end{eqnarray*}}

\newcommand{\cm}{\ensuremath{\mathcal M}}

\newcommand{\m}{\ensuremath{\mu}}
\newcommand{\n}{\ensuremath{\nu}}

\nopagebreak

\begin{document}
\baselineskip 22pt
\begin{flushright}
\end{flushright}
\vskip 65pt
\begin{center}
{\large \bf
Next-to-leading order QCD corrections to $W^+W^-$ production at the LHC 
in Randall Sundrum model 
}
\\
\vspace{8mm}
{\bf
Neelima Agarwal$^a$
\footnote{neel1dph@gmail.com},
}
{\bf
V. Ravindran$^b$
\footnote{ravindra@hri.res.in}, \\
Vivek Kumar Tiwari$^a$
\footnote{vivekkrt@gmail.com},
Anurag Tripathi$^c$
\footnote{anurag@theory.tifr.res.in}
}\\
\end{center}
\vspace{10pt}
\begin{flushleft}
{\it
a)~~Department of Physics, University of Allahabad, Allahabad 211002, India. \\
b)~~Regional Centre for Accelerator-based Particle Physics,\\
~~~~Harish-Chandra Research Institute, Allahabad 211019, India.\\
c)~~Department of Theoretical Physics, Tata Institute of Fundamental Research, \\ ~~~~Mumbai 400005, India. \\
}
\end{flushleft}
\vspace{10pt}
\centerline{\bf Abstract}
We present next-to-leading order QCD corrections to production of two $W$ bosons  
at the LHC in the Randall-Sundrum model.
Various kinematical distributions are obtained to order $\alpha_s$ in QCD by 
taking into account all the parton level subprocesses.  
We estimate the impact of the QCD corrections on various observables and find that they
are significant.  We also show the reduction in factorization scale uncertainty when
${\cal O}(\alpha_s)$ effects are included.

\vskip12pt
\vfill
\clearpage

\setcounter{page}{1}
\pagestyle{plain}

\newcommand{\vn}{\ensuremath{{\vec n}}}
The main aim of the upcoming Large Hadron Collider (LHC) is the search 
of the missing piece of the standard model (SM) ie. the Higgs boson 
and the existence of 
new physics which offers the solution to the hierarchy problem of SM.
In this direction, there exist many models based on ideas of
supersymmetry (SUSY), 
extra dimensions, technicolor etc. The possible existence
of new spatial dimensions beyond 3+1, came from early works of Kaluza and 
Klein in which they postulated a fifth dimension to unify electromagnetism
and gravity \cite{Kaluza:1921tu}, but the renaissance of extra dimensions 
began with the proposals of Arkani-Hamed, Dimopoulos and Dvali (ADD)
\cite{Antoniadis:1998ig}, and Randall and Sundrum (RS)\cite{Randall:1999ee}.

The RS model is a 5-dimensional theory with the fifth dimension 
compactified on an $S^1/Z_2$ orbifold
with a radius $R_c$. The Planck brane is located at the orbifold fixed point
$\phi=\pi$ while the SM fields are localized
at the  TeV brane which is at $\phi=0$. This geometry gives the following
metric in 5-dimensions:
\begin{equation}
ds^2 = e^{-2{\cal K}R_c|\phi|}\eta_{\mu\nu}dx^{\mu}dx^{\nu}~+~R_c^2d\phi^2  
\label{eq1}
\end{equation}
where $0 \leq \phi \leq \pi$.
To explain the hierarchy between the Planck scale and the electroweak (EW) scale 
we need ${\cal K}R_c$ only of the order ${\cal O}(10)$.
Introducing an extra scalar field in the bulk \cite{Goldberger:1999uk,Csaki:1999mp} 
showed that ${\cal K}R_c$ can be made stable against the 
quantum fluctuations. 

The variations of the above setup have also been considered in the literature
where the SM fields, except for the Higgs field, have been allowed to propagate
in the bulk \cite{Davoudiasl:1999tf, Grossman:1999ra, Gherghetta:2000qt}.
This framework provides an interesting new approach to the
flavor problem, as now also the hierarchical structures observed in the masses and
the mixing of the SM fermions could be explained in terms of geometrical effects  
\cite{Grossman:1999ra},
\cite{Gherghetta:2000qt ,ArkaniHamed:1999dc ,Huber:2000ie ,Huber:2003tu}. 
We will consider the original proposal of RS for our analysis.

The effect of extra dimensions on the SM fields 
is felt through the $KK$ gravitons. These $KK$ gravitons, $h^{({n})}_{\mu\nu}$, couple
to the SM energy momentum tensor and the interaction Lagrangian is
\begin{eqnarray}
{\cal L}_{int} & \sim & -{1\over {\overline{M_{Pl}}}}
T^{\mu\nu}(x) h^{(0)}_{\mu\nu}(x)
-{e^{\pi {\cal K} R_c} \over {\overline{M_{Pl}}}} \sum_{n=1}^{\infty}
T^{\mu\nu}(x) h^{(n)}_{\mu\nu}(x) \ .
\label{eq2}
\end{eqnarray}
$T^{\mu\nu}$ is the
symmetric energy-momentum tensor for the SM particles on the
3-brane, and ${\overline{M_{Pl}}}$ is the reduced Planck scale.
The masses of the $h^{({n})}_{\mu\nu}$ are given by
\begin{eqnarray}
M_n & = & x_n {\cal K} ~e^{-\pi {\cal K} R_c} \ ,
\label{eq3}
\end{eqnarray}
where the $x_n$ are the zeros of the Bessel function $J_1(x)$.
The first term in the interaction Lagrangian gives the coupling
of the zero-mode and it is Planck scale suppressed. The coupling
of the massive $KK$ states is enhanced due to the exponential factor
$e^{\pi {\cal K} R_c}$ and gives interactions of EW 
strength.
Consequently, except for the overall warp factor in the RS case,
the Feynman rules in the RS model are the same as those for the ADD case
\cite{Giudice:1998ck,Han:1998sg,Mathews:2004xp}.
The basic parameters of the RS model are
\begin{eqnarray}
m_0 & = & {\cal K} e^{-\pi {\cal K} R_c} \ , \nonumber \\
c_0 & = & {\cal K}/{\overline{M_{Pl}}} \ ,
\label{eq4}
\end{eqnarray}
where $m_0$ is a scale of the dimension of mass 
and $c_0$ ($0.01 \le c_0 \le 0.1$) is
an effective coupling. 
For our analysis we choose to work with the RS parameters $c_0$ and
$M_1$ the first excited mode of the graviton rather than $m_0$.

Summing  over all the $KK$ states we  obtain  effective graviton propagator : 
\begin{eqnarray}
{\cal D}_{Q^2} &=& \sum_{n=1}^\infty \frac{1}{Q^2 - M_n^2 + i M_n \Gamma_n}
\equiv {\lambda \over m_0^2} \ ,
\label{eq15}
\end{eqnarray}
where $M_n$ are the masses of the individual resonances (see Eq.~\ref{eq3}) and the $\Gamma_n$
are the corresponding widths.  

There are two ways to probe such effects at colliders, either through graviton emission 
or by virtual graviton exchange.  In this paper we will consider only virtual spin-2 $KK$ states. 
Production of boson pairs is one of the important process
at the LHC both in the context of SM and new physics
studies. Studies in other channels have been reported in  
\cite{Mathews:1999iw}
in extra  dimension models.
In this paper we will consider production of
$W$ pair at the LHC. Owing to its importance many studies have been 
carried out for its production in the SM; 
a study in the context  
of anomalous triple gauge boson vertices was carried out in  \cite{Hagiwara:1986vm,Baur:1995uv}.
Leading order (LO) studies in the SM can be found
in \cite{Brown:1978mq}. 
As is well known the LO results are highly sensitive to the arbitrary
renormalization and factorization scales. At this order the
factorization scale $\mu_F$ enters solely through the parton distribution
functions as the parton level cross-section, at this order,
does not depend on $\mu_F$.
As we include higher order terms of the perturbation series the dependence
will reduce and an all order result will be completely independent of these
arbitrary scales. In addition the NLO results are usually
significantly enhanced as compared to the LO results. It is thus important to carry out NLO calculation to
reduce these scale dependencies.
Because of its importance, its production has been studied to 
next-to-leading-order (NLO) accuracy
in the SM \cite{Ohnemus:1991kk,Frixione:1993yp,Ohnemus:1994ff,Dixon:1998py}.
These results were subsequently updated in \cite{Campbell:1999ah ,Dixon:1999di}.
These studies provide the precise estimate of higher order effects through $K$ factor 
as well as the sensitivity of the predictions to factorization scale.  
Its production has also been studied via gluon fusion through a quark box loop
or triangle quark loop with $\gamma$ or $Z$ boson exchange
\cite{Kao:1990tt}
and at one and two loop level in high energy limit in SM 
\cite{Chachamis:2008xu}.

Two $W$ bosons can couple to Kaluza Klein $(KK)$ gravitons, 
so it is possible to produce them  
through virtual graviton exchange at LO 
\cite{Agashe:2007zd}.
The significance of NLO computations in the extra dimension models for  
Drell-Yan \cite{Mathews:2004xp},
diphoton \cite{Kumar:2008pk},
$ZZ$ \cite{Agarwal:2009xr}, 
graviton+photon \cite{Gao:2009pn}, graviton+jet \cite{Karg:2009xk}
production has already been demonstrated.
Although NLO results are available in SM, they do not exist in literature in 
the context of RS model for $W$ boson pair production, 
which is the material of the present paper.

Before we present the results let us present in brief the pieces of
NLO calculation. The details can be found in 
\cite{Neelima} where we have given 
the matrix elements etc. for the process in the context of ADD model.
The signal comprises of contributions 
\be
|\cm_{SM}|^2 + |\cm_G|^2 +(\cm_{SM} \cm_G^* +c.c. )
\ee
where the first term is pure $SM$, the second is purely gravity
mediated and the third term is the interference of $SM$ and 
gravity mediated processes.
At leading order in strong coupling $\cm_{SM}$ has three contributions;
a t-channel or u-channel process and s-channel processes via $\gamma$ and $Z$ boson.
\be
q  \qb  \stackrel{t/u}{\rightarrow } W^+W^- ,  \quad \quad
q \qb \stackrel{s, \gamma} {\rightarrow} W^+W^- , \quad \quad
q \qb \stackrel{s, Z} {\rightarrow} W^+W^-
\ee 
As the $KK$ gravitons couple with same strength to quarks
and gluons  both quark and gluon initiated Feynman diagrams with
s-channel graviton propagator contribute to $\cm_G$.

Next at order $\alpha_s$ we have to include both one loop corrections
to the above processes and also real emission contributions in which
in addition to $W^+W^-$ a parton is emitted in the final state.
The soft and collinear configurations in the loop integrals give
divergences which we have regulated using dimensional regularization 
$(n=4+\eps)$
thus the singularities appear as simple and double poles in $\epsilon$.
As the process under consideration is $UV$ finite, these poles are
only soft and collinear.  
In the real emission case we have $q \qb, ~ qg$ and $gg$ initiated 
processes. As we have $gg \rightarrow W^+W^-$ at leading order 
through graviton exchange, we note that all the 4-kinds of splitting
functions $P_{qq}, P_{qg}, P_{gq}, P_{gg}$ are involved in the
calculation. In addition to the above soft and collinear singularities,
the other set of these divergences appear from phase space integration 
of the real emission matrix elements. The sum of virtual and real 
contributions is completely finite ie. free of poles in $\epsilon$
after mass factorization is carried out. We have used $\overline {MS}$ scheme throughout,
both for the renormalization and factorization.

We have employed the method of two cutoff phase space slicing to handle the 
real emission processes. In this method two small dimensionless slicing
parameters $\delta_s$ and $\delta_c$ are introduced to 
divide the real emission phase space into soft and collinear regions (for
a review of the method please see  \cite{Harris:2001sx}). 
The cross section can be written as, then,
\be
d \sigma = d \sigma^{LO} + d \sigma^{\rm virt} 
           + d \sigma^{\rm soft + col + CT} (\delta_s,\delta_c)
           + d \sigma^{\rm hard~ non ~col} (\delta_s,\delta_c)
\ee
Here the third term gives the contribution coming from 
the soft and collinear regions which is rendered finite
after adding the counter term (CT) 
for mass factorization. The last term denotes the contribution 
of hard non collinear configurations and is finite.
We define 
\bea 
d \sigma ^{\rm 2-body} 
= d \sigma^{\rm virt} 
+ d \sigma^{\rm soft +col +CT} (\delta_s, \delta_c)
\\
d \sigma^{\rm 3-body}
= d \sigma^{hard~non~col}( \delta_s, \delta_c)
\eea
Note that, individually $d \sigma^{2-body}$ and $ d \sigma^{3-body}$
depend on $\delta_s$ and $\delta_c$ but the sum should be independent
of the parameters which were introduced to slice the phase space.
We have incorporated all the above details in our monte carlo code 
which is implemented on FORTRAN 77 and easy to tailor for various 
cuts on the final state bosons.

We now make some general remarks about the computation.
We have used Feynman gauge in QCD sector and unitary gauge in 
electroweak sector. The choice of unitary gauge simplifies the
calculation as both the electroweak Goldstone bosons and ghosts disappear.
Further we note that the term proportional to $1/\xi$ in gluon-gluon-graviton
vertex does not contribute.  Also the results do no depend on the 
arbitrary vector $n^{\mu}$ which appears in gluon polarization sum:
\be
\epsilon^{\mu} (k) \epsilon^{\nu *}  (k)= -g^{\mu \nu} + \frac{k^\m n^\n + k^\n n^\m}{k \cdot n}
\ee
Further our SM matrix elements agree with 
those given in \cite{Ohnemus:1991kk,Frixione:1993yp}. 
To check the numerical implementation of the phase space slicing method
we have checked the stability of the sum of $2-body$ and $3-body$ contributions
against variation of slicing parameters $\delta_s$ and $\delta_c$ and we found
the sum to be stable over a wide range of these parameters. In what follows we will
use $\delta_s=10^{-3}$ and $\delta_c=10^{-5}$.


We now present the kinematical distributions for the $W^+W^-$ production at the LHC. 
The LHC with a center of mass energy of $14~TeV$ will be our 
default choice. However we will also present some results for a center of mass energy of $10~TeV$ for the LHC.
For numerical evaluation, the following SM parameters \cite{Amsler:2008zzb} will be used
\be 
m_W= 80.398~ GeV,  \quad m_Z = 91.1876~ GeV,\quad \Gamma_Z=2.4952 ~GeV, \quad \sin^2 \theta_W = 0.231
\ee
where $\theta_W$ is the weak mixing angle.
For the electromagnetic coupling constant $\alpha$ we use $ \alpha^{-1} = 128.89$. 
CTEQ6 \cite{Pumplin:2002vw} density sets are used for parton distribution 
functions. 2-loop running for the strong coupling constant is used .
The number of active light-quark flavors is taken to be 5 and the value of $\Lambda_{QCD}$ is 
chosen as prescribed by the CTEQ6 density sets. At leading order  we
use CTEQ6L1 density set ( which uses the LO running $\alpha_s$ ) with the corresponding 
$\Lambda_{QCD}=165~MeV $. At NLO we use CTEQ6M density set ( which uses 2-loop running $\alpha_s$ )
with the $\Lambda_{QCD}=226~MeV $; this value of $\Lambda_{QCD}$ enters into the evaluation of the 
2-loop strong coupling.
The  default choice for the renormalization and factorization scale is the identification
to the invariant mass of the $W$ boson pair ie., $\mu_F =\mu_R =Q$. Furthermore the
$W$ bosons will be constrained to satisfy $|y_W| < 2.5$, where $y_W$ is 
the rapidity of a final state $W$ boson .

We present invariant mass ($Q$) and rapidity ($Y$) distribution of 
the $W$ boson pairs. These kinematical variables are defined as
\be
Q^2 = (p_{W^+} +p_{W^-})^2, \quad \quad Y= \frac{1}{2} \ln \frac{P_1 \cdot q}{P_2 \cdot q},
\ee
where $P_1$ and $P_2$ are the momenta of colliding hadrons, and  $q=p_{W^+} + p_{W^-}$
denotes the sum of the $W$-boson 4-momenta. In obtaining these distributions
all order $\alpha_s$ contributions have been taken into account. 

\begin{figure}[ht]
\centerline{\epsfig{file=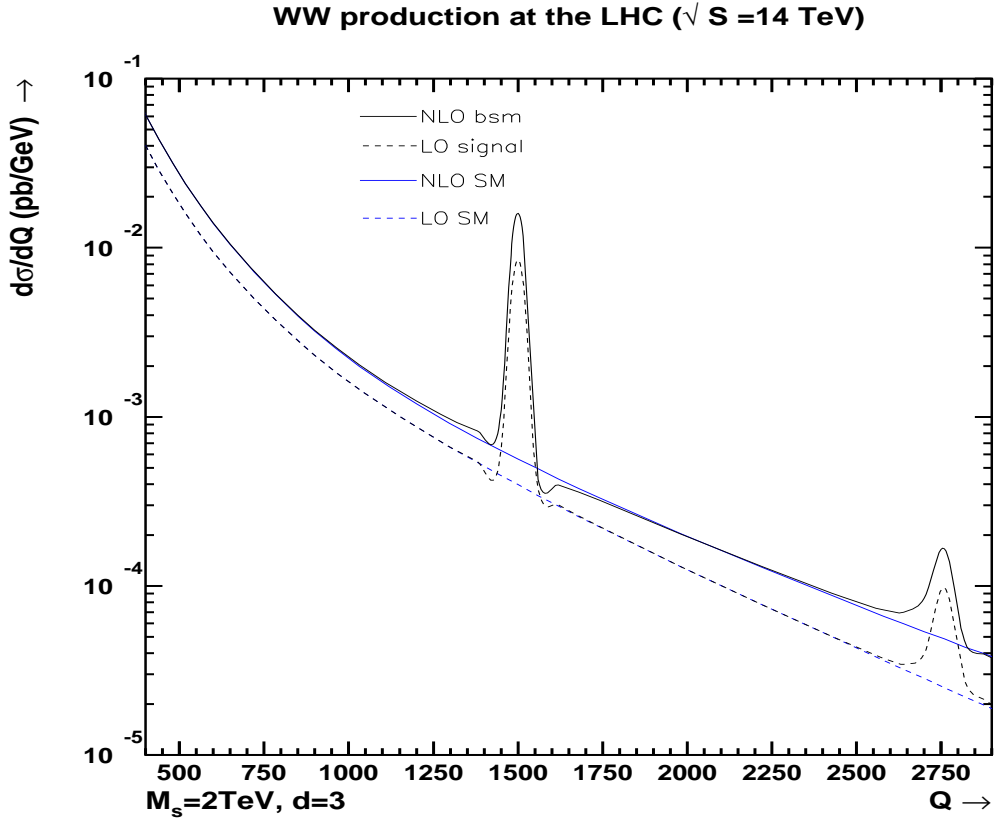,width=15cm,height=12cm,angle=0} }
\caption{Invariant mass distribution for SM and signal both at LO
and NLO. Dash-dot curves represent LO results and solid curves give
NLO results. We have chosen $M_1 =1500~GeV$ and the parameter $c_0 =0.01$.  }
\label{inv}
\end{figure}

In Fig.~\ref{inv} we have plotted the invariant mass distribution both for the SM and the 
signal for LHC at $14 TeV$.
The two curves with peaks correspond to the signal and the remaining two curves give SM
predictions.  Here we have chosen $c_0 =0.01$ and $M_1 =1500 GeV$.
To highlight the importance of QCD corrections we have also displayed 
the LO results of SM and the signal, and 
we observe that at $Q=1500~GeV$ the $K$ factors (defined as $K=d\sigma^{NLO}/d\sigma^{LO}$) has a value 1.9. Thus NLO QCD corrections give a substantial
enhancement over the LO predictions.

\begin{figure}[ht]
\centerline{\epsfig{file=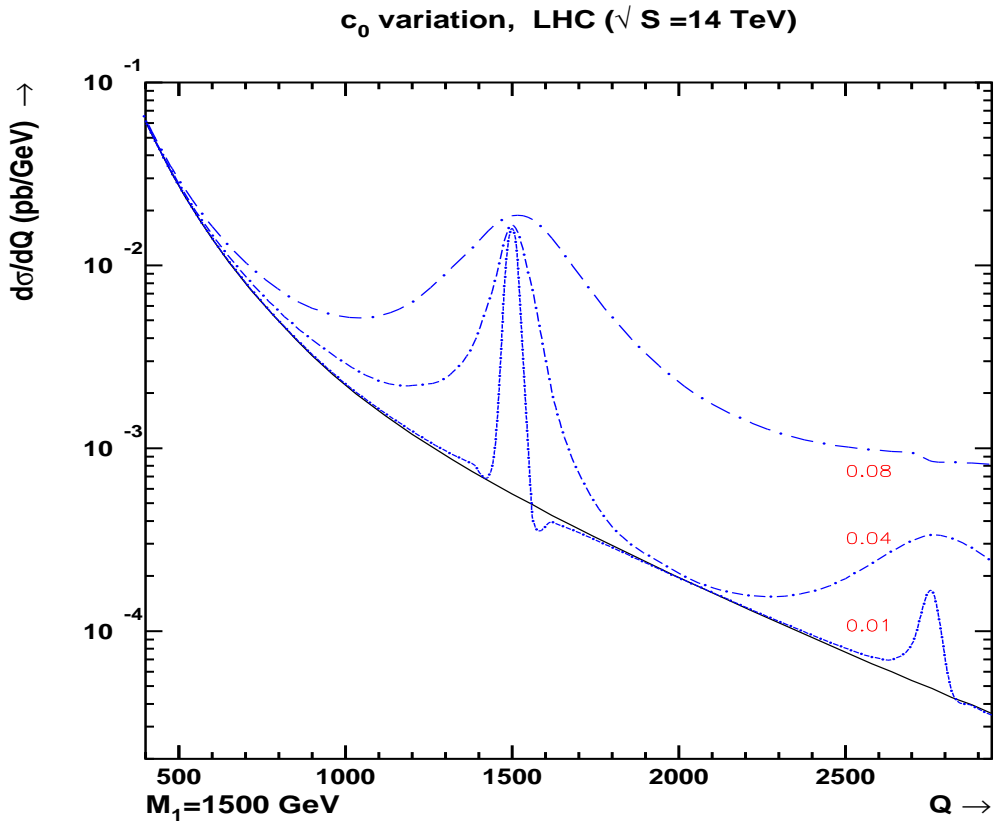,width=15cm,height=12cm,angle=0} }
\caption{
Effect of variation of $c_0$ on invariant mass distribution. 
All the curves correspond to NLO results with $M_1$ fixed
at $1500~GeV$. The solid curve corresponds to SM and the dash-dot curves
to the signal. The signal is plotted for $c_0 =0.01, 0.04, 0.08$ and 
the dash size increases with increasing $c_0$ }
\label{cvar}
\end{figure}

Next we present in Fig.~\ref{cvar} the effects of varying the parameter $c_0$ on the
invariant mass distribution. All the curves shown correspond to NLO results, and
we have also plotted the SM background for comparison.

\begin{figure}[ht]
\centerline{\epsfig{file=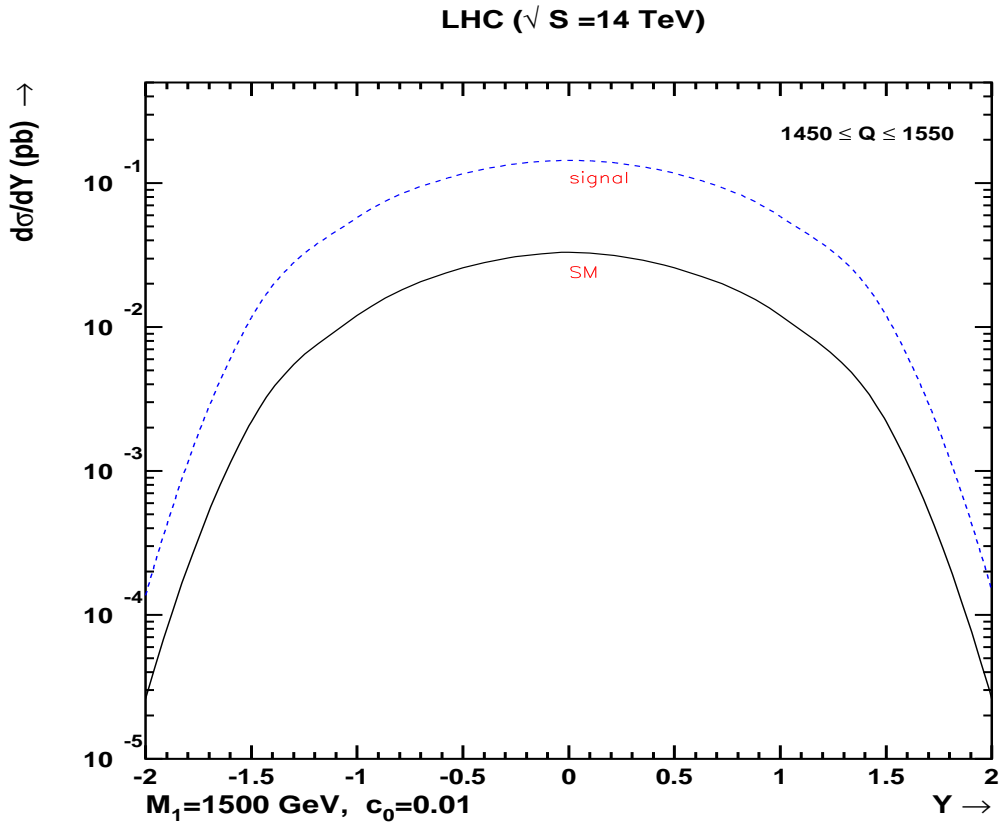,width=15cm,height=12cm,angle=0} }
\caption{Rapidity distribution for SM and signal at
NLO. Dash curve represents the signal and solid curve gives SM result.
 We have chosen $M_1 =1500~GeV$ and the parameter $c_0 =0.01$.
To enhance the signal we have integrated over $Q$ in the range $1450 \leq Q \leq 1550$.  }
\label{y}
\end{figure}

In Fig.~\ref{y} we have plotted the rapidity distribution $d\sigma/dY $ at 
NLO both for SM and the signal for $c_0=0.01$. 
We have plotted this distribution in the interval $-2.0 < Y < 2.0 $ and 
have carried out an integration over the invariant mass interval $1450 < Q < 1550$
to increase the signal over the SM background.
As expected the distribution is symmetric about $Y=0$.

\begin{figure}[ht]
\centerline{\epsfig{file=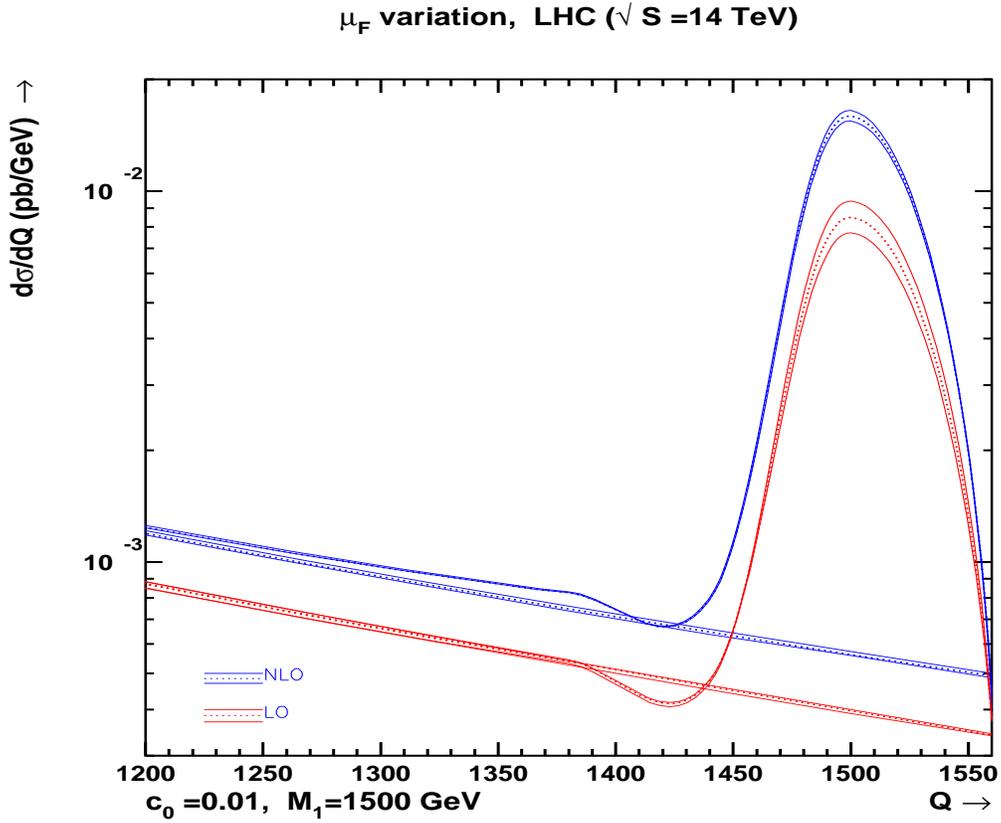 ,width=15cm,height=12cm,angle=0} }
\caption{Factorization scale variation in the invariant mass distribution.
The curves correspond to $c_0 =0.01$ and $M_1 =1500 GeV$ at the LHC at $\sqrt{S}=14~ TeV$.
The $\mu_F$ is varied between $Q/2$ and $2Q$. The dash curves correspond to $\mu_F=Q$  }
\label{mufvar}
\end{figure}

As was noted above the NLO QCD corrections reduce the sensitivity 
of the cross sections to the factorization scale $\mu_F$; this we now show in 
the Fig.~\ref{mufvar}. We have plotted SM and the signal both at LO and NLO,
and have varied the factorization scale $\mu_F$ in the range $Q/2 < \mu_F < 2Q$.
The central curve in a given band (shown by the dotted curves) correspond to 
$\mu_F =Q$. In all these results the renormalization scale is fixed at $\mu_R =Q$.
We notice that the factorization scale uncertainty at  LO 
is 21.8 \% at $Q=1500~GeV$ as compared to 6.7 \% at NLO. Thus we see that
NLO computation achieves significant reduction in uncertainty and makes
predictions much more precise.

\begin{figure}[ht]
\centerline{\epsfig{file=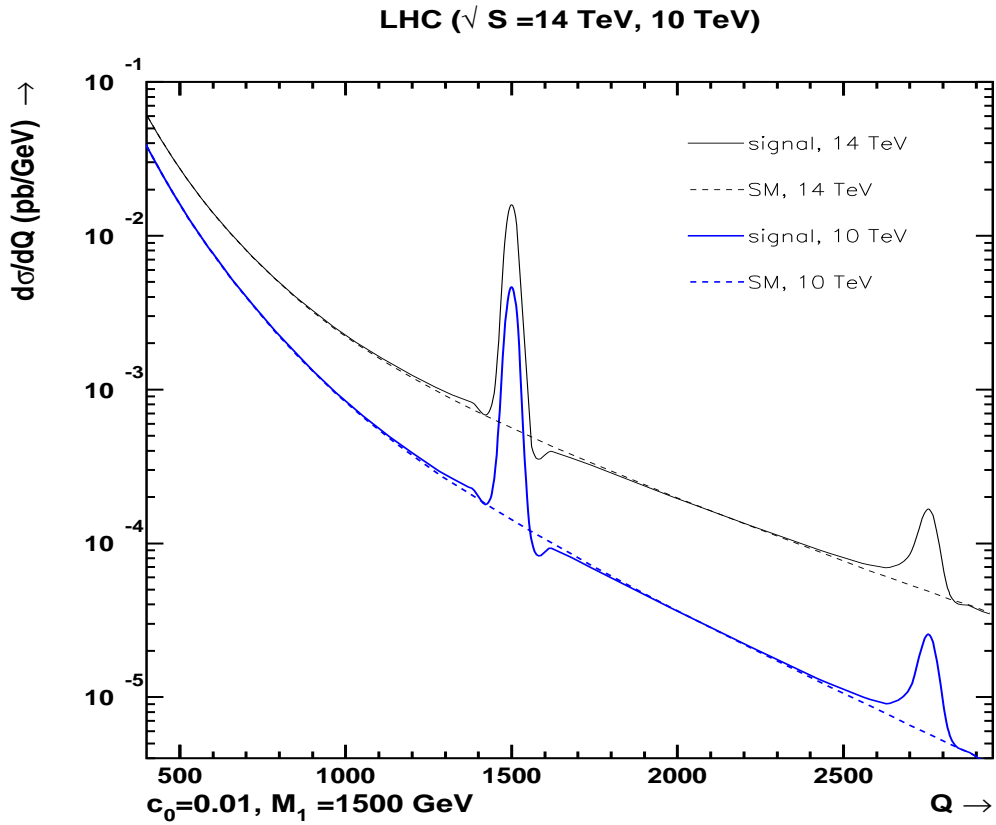,width=15cm,height=12cm,angle=0} }
\caption{Invariant mass distribution for SM and signal at $ \sqrt{S} =10 TeV$ and
$14 TeV$. All the curves correspond to NLO results. We have chosen $M_1 =1500~GeV$ and the parameter $c_0 =0.01$.  }
\label{ten}
\end{figure}

At the end we present in Fig.~\ref{ten}, $d\sigma/dQ$ for LHC with a centre of mass
energy of $10~TeV$ at NLO both for SM and signal. For comparison 
we have also plotted the $14~TeV$ results in the same figure.


To summarize, in this paper we have carried out a full NLO QCD calculation for 
the production of two $W$ bosons at the LHC at $14~TeV$ and $10~TeV$ in the
extra dimension model of Randall and Sundrum. Here we take all order $\alpha_s$ contributions,
both in the SM and in the gravity mediated processes and their interferences,
into account. 
We have presented invariant mass and rapidity distributions both at LO and NLO. 
We use CTEQ 6L1 and CTEQ 6M parton density sets for LO and NLO observables, respectively.
Significant enhancements over the LO predictions are observed.
The $K$ factor are large and at $Q=1500~GeV$ (we have taken this as the first RS resonance) 
$K=1.9$. This justifies the entire exercise of carrying out a NLO computation.
The effect of variation of parameter $c_0$ in
invariant mass distribution is also presented. We have shown that a significant
reduction in LO theoretical uncertainty, arising from the factorization scale, is achieved 
by our NLO computation. It is observed that an uncertainty of 21.8 \% at LO 
as $\mu_F$ is varied between $Q/2$ and $2Q$ is reduced to
6.7 \%. 
Thus our NLO results are more precise than the LO results
and suitable for further studies for constraining the parameters of the RS model.
Invariant mass distribution is also presented for LHC at a center of mass energy of
$10 TeV$ at the NLO level.
\\

\noindent
{\bf Acknowledgments:}
The work of NA is supported by CSIR Senior Research Fellowship, New Delhi.
NA, AT and VR would also like to thank
the cluster computing facility at Harish-Chandra Research Institute. 
NA and VKT acknowledge the computational support of the computing facility which has been developed by
the Nuclear Particle Physics Group of the Physics Department, Allahabad University under the Center of
Advanced Study (CAS) funding of U.G.C. India. The authors would like to thank Prakash Mathews 
and M.C. Kumar for useful discussions.



\end{document}